\documentclass[
superscriptaddress,
preprint,
graphicx,
floatfix,
bibnotes,
amsmath,
amssymb,
aip,
]{revtex4-2}

\usepackage{graphicx}% Include figure files
\usepackage{dcolumn}% Align table columns on decimal point
\usepackage{bm}% bold math

\usepackage{mathtools}
\usepackage{color}

\begin{document}

\title{Spectroscopic signatures of mode-selective electron-phonon coupling in transient reflectivity change on charge-density-wave TiSe$_2$}

\author{Yu Mizukoshi}
\email{Authors to whom correspondence should be addressed: mizukoshi.yu.tkb\_gf@u.tsukuba.ac.jp}
\affiliation{Department of Applied Physics, Graduate school of Pure and Applied Sciences, University of Tsukuba, 1-1-1 Tennodai, Tsukuba 305-8573, Japan}
\author{Muneaki Hase}
\email{mhase@bk.tsukuba.ac.jp}
\affiliation{Department of Applied Physics, Graduate school of Pure and Applied Sciences, University of Tsukuba, 1-1-1 Tennodai, Tsukuba 305-8573, Japan}

\date{\today}% It is always \today, today,
             %  but any date may be explicitly specified

\begin{abstract}
We report on time- and spectrally resolved pump–probe spectroscopy measurement on the charge-density wave (CDW) state of TiSe$_2$ at helium temperatures. The non-oscillatory component of the spectrally resolved signals exhibit a sign reversal near 1.5 eV, which was interpreted as a redshift of the optical transition energy originated from the interband transition between the Se $p$-based valence band to the Ti $d$-derived conduction band. Furthermore, the oscillatory components exhibit distinct spectral dependences between the optical phonons and the CDW-derived modes, suggesting the mode-selective coupling.
\end{abstract}

\maketitle
\newpage
%\section{Introduction}
A charge-density wave (CDW) is a macroscopic quantum state characterized by a periodic modulation of the electronic charge density accompanied by a periodic lattice distortion \cite{balandin2021charge,gruner1988dynamics}. One distinctive feature of this quantum state is some characteristic properties due to collective dynamics of condensed electrons  such as nonlinear transport, frequency-dependent conductivity, and current oscillations \cite{gruner1988dynamics}, opening new applications opportunities \cite{balandin2021charge}. At the same time, many physical questions in CDW physics remain unresolved, including the mechanism of CDW formation \cite{zhu2017misconceptions}, continuing to motivate fundamental interest. Toward microscopic understanding and ultrafast control of the CDW properties, femtosecond spectroscopy has opened routes to explore their non-equilibrium dynamics. Indeed, optical perturbation %to CDW stability 
has been shown to induce the melting of CDW order \cite{schmitt2008transient,eichberger2010snapshots} and the formation of hidden CDW phases \cite{kogar2020light,stojchevska2014ultrafast}, as well as ultrafast switching of resistance \cite{stojchevska2014ultrafast,vaskivskyi2016fast} and depinning \cite{jacques2016laser}.

TiSe$_2$ is a type of transition-metal dichalcogenides and has been widely discussed in its CDW phase. TiSe$_2$ forms a three-dimensional CDW below $T_c\sim 200$ K and exhibits several unusual properties such as large fluctuations \cite{monney2016revealing,fragkos2026electron,cheng2022light}, stability due to excitonic correlation \cite{cercellier2007evidence,kogar2017signatures}, possibility of chiral CDW \cite{Xu2020,nie2023unraveling}, photoinduced nonthermal melting \cite{rohwer2011collapse,mohr2011nonthermal,porer2014non}, and dimension crossover \cite{duan2021optical,cheng2022light}. Previous studies have shown that these rich properties are closely intertwined, and various measurements have provided complementary insights. From the viewpoint of optical pump-probe studies, coherent oscillations of the CDW amplitude mode (AM) have been discussed, for example in relation to nonthermal melting \cite{mohr2011nonthermal,hedayat2019excitonic,hedayat2021investigation}. However, the microscopic details of how these coherent oscillations and photoexcited carriers couple to the optical transition remain unclear. To this end, probe-wavelength-dependent measurement is effective \cite{sayers2022spectrally}. In particular, the strong optical transition % from the $p$-based band to the Ti $d$-derived band 
near 800 nm (1.55 eV) \cite{bayliss1985reflectivity,buslaps1993spectroscopic}, the most commonly used in ultrafast measurements, should provide an important clue to understanding these pictures.

In this study, we performed time- and spectrally resolved pump–probe spectroscopy on the CDW state of TiSe$_2$ at helium temperatures. As the probe photon energy was varied from 1.40 to 1.58 eV, the $\Delta R/R$ signal exhibited an evolution from a positive to a negative value. It can be interpreted as the shifts of the optical transition energy due to photoexcited carriers and lattice heating. In the coherent oscillation part, four modes were observed, consisting of one optical phonon and three CDW-derived modes. Interestingly, the amplitudes of the CDW-derived modes exhibited a probe-energy dependence distinct from that of the optical phonon, suggesting the mode-selective coupling. Our results provide an important route toward understanding how photoexcited carriers and coherently excited phonons couple to the optical transition in CDW systems.

%\section{Experimental setup
Spectrally resolved optical pump-probe measurements were performed using a Ti:Sapphire oscillator and a home-built monochromator, as illustrated schematically in Fig. 1(a). The femtosecond oscillator had a central wavelength of 825 nm, a repetition rate of 80 MHz, a pulse duration of 35 fs and a spectral bandwidth of 50 nm of a full width at half maximum (FWHM). To observe the transient reflectivity change ($\Delta R/R$) with high singal-to-noise ratio, a fast-scan technique was employed, in which the pump path was modulated at 19.5 Hz. The spot size of the pump and probe was set at $\sim$60 \textmu m and $\sim$30 \textmu m, respectively, while the absorbed fluence of the pump pulse was fixed at 3.45 \textmu J/cm$^2$, which is below the CDW melting threshold. The reflected probe pulse was spectrally selected by a monochromator and detected using a balanced detection with Si-PIN photodiodes. The monochromator consists of a reflective-type diffraction grating and a pair of mechanical slits. The widths of both the entrance and exit slits were set to 500 \textmu m, yielding a wavelength resolution with FWHM of less than 7 nm. The monochromator was calibrated with a commercial spectrometer. All measurements were carried out at 4.8 K under high-vacuum conditions using a closed-cycle cryostat. The cryogenic environment provides the advantages of a large coherent phonon amplitude and a stable CDW state.

The bulk single crystal TiSe$_2$ was provided by HQ graphene. For pump-probe measurements, TiSe$_2$ sample was freshly cleaved before being mounted in a chamber. The equilibrium optical reflectance of TiSe$_2$ was estimated from the Drude–Lorentz model reported in previous studies \cite{li2007semimetal,tyulnev2025high}.  The complex refractive index of our sample was also analyzed using spectroscopic ellipsometry (UVISEL PLUS, Horiba) at ambient conditions. The reflectance spectra obtained from the literature values and from our measurements are shown in Fig. 1(b), showing qualitative agreement. In the probe energy range used in this study, 1.40–1.58 eV, the reflectance exhibits a broad maximum governed by a optical transition at approximately 1.5–1.7 eV. This transition is considered to originate from the interband transition between the Se $p$-based valence band and the Ti $d$-derived conduction band near the $M$ points of the Brillouin zone \cite{bayliss1985reflectivity,buslaps1993spectroscopic,novko2025excitons}, and is not affected by the formation of the CDW \cite{novko2025excitons}. In the following, this optical transition is referred to as the $M$-point optical transition.

%\section{Results and Discussions}
The $\Delta R/R$ signal was first characterized without resolving the probe spectrum. Figure \ref{figure2}(a) shows the spectrally integrated $\Delta R/R$ signal. After exhibiting a negative drop around time zero, the signal turns positive within 200 fs and subsequently decays accompanied by coherent oscillations. This behavior deviates from the degenerate pump–probe report at 800 nm, in which the negative response decays without turning positive \cite{mohr2011nonthermal}. This discrepancy appears despite a wavelength difference of only 35 nm (65 meV), suggesting that the response exhibits a strong wavelength dependence around 800 nm.

The non-oscillatory component was extracted using bi-exponential decay functions with an offset term; $f(t) = A_0 + \sum_{i=1,2} A_i e^{-t/\tau_i}$. Here, $A_0$ denotes an amplitude of offset, $ A_i$ and $\tau_i$ represent the amplitude and lifetime of the $i$-th exponential component, respectively. The fitting was performed at $t > 350$ fs, where the signal begins to decrease. The fitting curve is shown as a blue line in Fig. 2(a), and the corresponding residual is displayed in Fig. 2(b). The residual contains only the oscillatory components, confirming the validity of the fitting.
The extracted lifetimes of the exponential components were $\tau_1$ = 608 fs and $\tau_2$ = 6.0 ps, which are qualitatively consistent with the lifetimes of the photoexcited carriers reported by time- and angle-resolved photoemission spectroscopy (tr-ARPES) under the midinfrared photon excitations \cite{monney2016revealing}.

The oscillatory component in Fig. 2(b) can be attributed to coherently excited optical phonons and CDW collective excitations. Figure 2(c) shows the fast Fourier transform (FFT) spectrum, revealing that the coherent response consists of some vibration modes. %3.5 THz, 5.2 THz, and 6.2 THz peaks. 
According to Raman scattering\cite{holy1977raman,sugai1980raman} and temperature dependence shown in supplemental materials with references \cite{sugai1980raman,sutar2024photo,snow2003quantum,holy1977raman}, these vibrations can be assigned to the 3.4 THz zone-folded phonon (ZFP$_1$), 3.5 THz amplitude mode (AM), 5.2 THz zone-folded phonon (ZFP$_2$) and the 6.2 THz A$_{1g}$ optical phonon, respectively. Although ZFP$_1$ and AM have almost degenerate frequencies, the AM has a larger amplitude and a shorter lifetime. Therefore, a sharp ZFP$_1$-derived dip was observed within the AM-derived main peak in the FFT spectrum. The contribution of ZFP$_1$ can be isolated by extracting the signal after 10 ps, shown as a blue spectrum in Fig. 2(c). Our measurements clearly identified these four modes, which had been difficult to resolve in previous studies \cite{mohr2011nonthermal,hedayat2019excitonic,hedayat2021investigation}.

The reflected probe was then spectrally resolved. Figure 3(a) shows a color map of the spectrally resolved $\Delta R/R$ signals, with representative temporal traces at selected probe wavelengths shown in Fig. 3(b). The signal is found to be positive on the lower-energy side of the probe spectrum, whereas it turns negative on the higher-energy side across approximately 1.52 eV. %The behavior at 1.55 eV (800 nm) was almost consistent with that reported in previous study \cite{mohr2011nonthermal}. 
Although the peak times of the signals were also found to vary with the probe wavelength, this variation is likely caused by a wavelength-dependent initial response, as discussed in the Supplemental materials with reference \cite{kirby2020transient}.

To quantify the amplitudes and lifetimes of the non-oscillatory components, the time-domain response was fitted by $f(t)$ for each wavelength at $t > 350$ fs. The fitting results are displayed as gray lines in Fig. 3(b), and the obtained parameters are summarized in Figs. 4(a)-(c). Both the amplitudes of the offset $A_0$ and the exponential components $A_1, A_2$ exhibited the same trend, reaching maxima around 1.45 eV and exhibiting a sign reversal around 1.52 eV. This common trend suggests that the three components modulate the dielectric function with a similar spectral profile. In addition, the extracted lifetimes were almost independent of the probe wavelength and were consistent with the lifetimes of the photoexcited carriers \cite{monney2016revealing}.

The sign reversal observed around 1.52 eV can be explained by a redshift of the $M$-point optical transition energy. An increase in electronic temperature or carrier density induced by photoexcitation can lead to a reduction in the optical transition energy \cite{heinrich2023electronic}, which is known as band-gap renormalization in semiconductors \cite{fukuda2024coherent}. In this case, $\Delta R/R$ exhibits a derivative feature of the reflectance spectrum \cite{heinrich2023electronic}. Therefore, a sign reversal is expected near the reflectance extrema, schematically illustrated in Fig. 4(d). In the present study, the probe wavelength is located near a local maximum of the reflectance, and this picture qualitatively accounts for the experimental observations.

The two-step relaxation processes represented by $\tau_1$ and $\tau_2$ reflect different carrier relaxation. After the non-equilibrium carriers thermalize into a hot Fermi–Dirac distribution through electron–electron scattering within 200 fs \cite{mathias2016self}, the elevated electronic temperature relaxes through electron–phonon scattering. %, described by the three-temperature model \cite{heinrich2023electronic}. The 
Subsequently, the electron-hole pairs undergo recombination, including Auger or phonon-assisted recombination \cite{dai2015ultrafast}. In semimetals, the time constants of electron–phonon scattering and electron–hole recombination have been reported to be on the sub-picosecond and several-picosecond timescales, respectively \cite{dai2015ultrafast}, supporting their assignment to $\tau_1$ and $\tau_2$. The offset term represents a long-lived contribution that barely relaxes in the scan range of approximately 25 ps, and can be associated with lattice heating. An increase in lattice temperature is also known to induce a redshift of optical transitions through thermal expansion or electron–phonon coupling \cite{heinrich2023electronic}. This is consistent with the experimental observation that the offset term exhibits a spectral modulation similar to the photoexcited-carrier response.

Then, to focus on the coherent response, FFT spectrogram of the oscillatory component was shown in Fig. 5(a). The FFT intensities were found to depend strongly on the probe wavelength. For a quantitative evaluation, the FFT intensities of the individual modes were extracted, as shown in Figs. 5(b)–5(e). To distinguish the signal of each mode from the spectral tail of the AM, the background level for each mode is also indicated by the gray shaded regions. All CDW-derived modes, AM, ZFP$_1$, and ZFP$_2$, show maximum FFT intensity around 1.48 eV that is clearly distinguishable from the background. By contrast, the $A_{1g}$ phonon shows larger intensity at the spectral tails, whereas its intensity decreases to nearly the noise floor near the 1.5 eV. This discrepancy imply that the CDW-derived modes and the $A_{1g}$ optical phonon have different coupling strengths to $M$-point optical transition.

The spectral dependence of the A$_{1g}$ phonon can be naturally understood in terms of the coupling to the $M$-point optical transition. Coherent phonons modulate the electronic bands or the optical transition energy through electron–phonon coupling such as deformation-potential interaction \cite{leuenberger2015classification,gerber2017femtosecond,fukuda2024coherent}. In this case, the coherent phonon amplitude is expected to exhibit a derivative feature \cite{sayers2022spectrally,heinrich2023electronic}, and the phonon signal should be strongly suppressed near a local maximum of the reflectance spectrum, as in carrier response. The reduction of the phonon intensity around 1.48 eV is consistent with the reduction of the carrier response amplitude at nearly the same energy, strongly supporting this interpretation. In other words, the A$_{1g}$ phonon, photoexcited carriers, and the lattice-heating affect the $M$-point optical transition energy.

It is interesting that all CDW-derived modes show the wavelength dependence distinct from that of the A$_{1g}$ phonon. This discrepancy suggests that the CDW modes are not strongly coupled to the $M$-point optical transition. Rather, CDW modes may instead strongly couple to optical transitions associated with the CDW phase transition. In principle, the CDW AM directly modulates the order parameter, namely the CDW gap \cite{gruner1988dynamics,leuenberger2015classification}. Therefore, the AM oscillation should strongly modulate optical transitions from the valence bands that form the CDW gap. Alternatively, AM may modulate the transitions that are sensitive to changes in the CDW gap. Although zone-folded phonons do not necessarily couple directly to the order parameter, their spectral fingerprints similar to those of the AM suggest strong coupling to similar optical transitions. However, because of Van Hove singularity of the Se-$p$ states \cite{novko2025excitons}, the reflectance in the 1–2.5 eV range is dominated by $M$-point optical transition in both the semimetallic and CDW phases, as shown in Fig. 1(b). Therefore, it is difficult to isolate contributions from other transitions. Further theoretical calculations to extract the relevant optical transition energies will be required to clarify the detailed coupling between coherent phonon and optical transition.

%\section{Conclusion}
In conclusion, we investigated the electronic and coherent vibrational responses in the CDW state of 1T-TiSe$_2$ at 4.8 K using time- and spectrally resolved pump–probe spectroscopy. The non-oscillatory components consist of photoexcited-carrier and lattice heating show a sign reversal around 1.5 eV, which was interpreted as a redshift of the optical transition energy from the Se $p$-derived band to the Ti $d$-derived band. In the coherent response, different probe-energy dependences were observed between the three CDW-derived modes and the A$_\text{1g}$ optical phonons, suggesting different coupling strengths or coupling mechanisms. These findings identify key phenomena relevant to ultrafast optical control in two-dimensional layered materials and CDW systems.

\section*{DATA AVAILABILITY}
The data that supports the findings of this study are available from the corresponding author upon a reasonable request.

\section*{Acknowledgement}
This work was supported by JSPS KAKENHI (Grant Numbers. 25KJ0687 and 22H01151). Y.M. acknowledges the support from JST SPRING, Japan Grant No. JPMJSP2124. M.H. acknowledges the support from the DGIST R\&D program (24-KUJoint-06 and 25-IRJoint-04).

\section*{AUTHOR DECLARATIONS}
\section*{Conflict of Interest}
The authors have no conflicts to disclose.

\section*{Author Contributions}
\textbf{Yu Mizukoshi:} Conceptualization (lead); Investigation (lead); Formal Analysis (lead); Writing-Original Draft Preparation (lead); 
%Wiring-review $\&$ editing (lead); 
Resources (supporting); Funding acquisition (supporting).
\textbf{Muneaki Hase:} Writing-Original Draft Preparation (equal); %Wiring-review $\&$ editing (equal);
Resources (lead); Funding acquisition (lead); Supervision (lead).

\bibliography{reference}% Produces the bibliography via BibTeX.

\begin{figure}[p]
    \centering
    \includegraphics[width = 8cm]{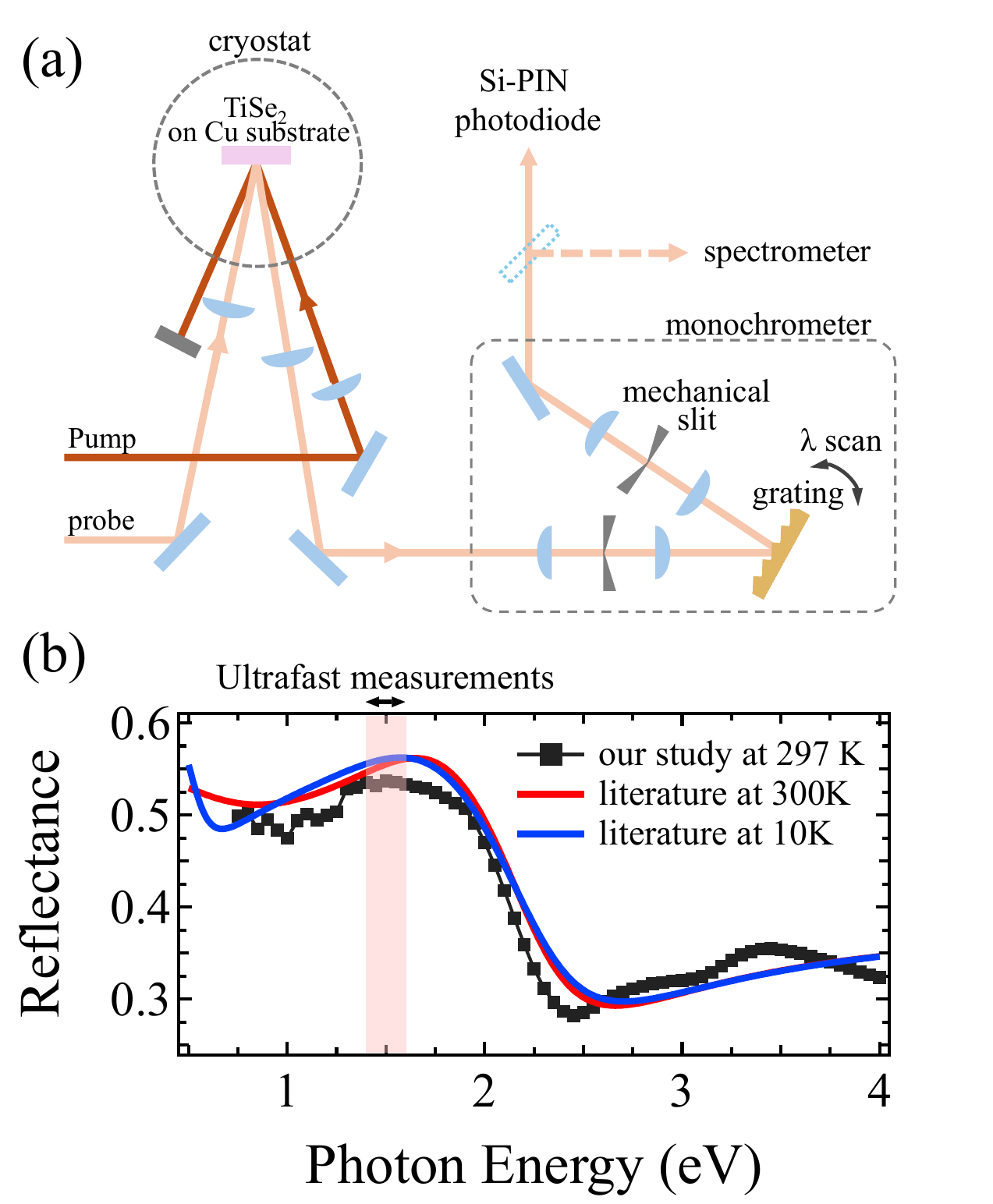}
    \caption{(a) Schematic illustration of the experimental setup. The pump and probe beams were focused using separate lenses, and the reflected probe beam was spectrally resolved by a home-built monochromator. The monochromator consisted of a diffraction grating and mechanical slits. (b) Optical reflectance spectrum of TiSe$_2$ . The black dots were calculated from ellipsometry measurements on our sample, while the red and blue curves correspond to calculations based on Drude-Lorentz parameters \cite{li2007semimetal,tyulnev2025high}.
    }
    \label{figure1}
\end{figure}
\clearpage

\begin{figure}[p]
    \centering
    \includegraphics[width = 8cm]{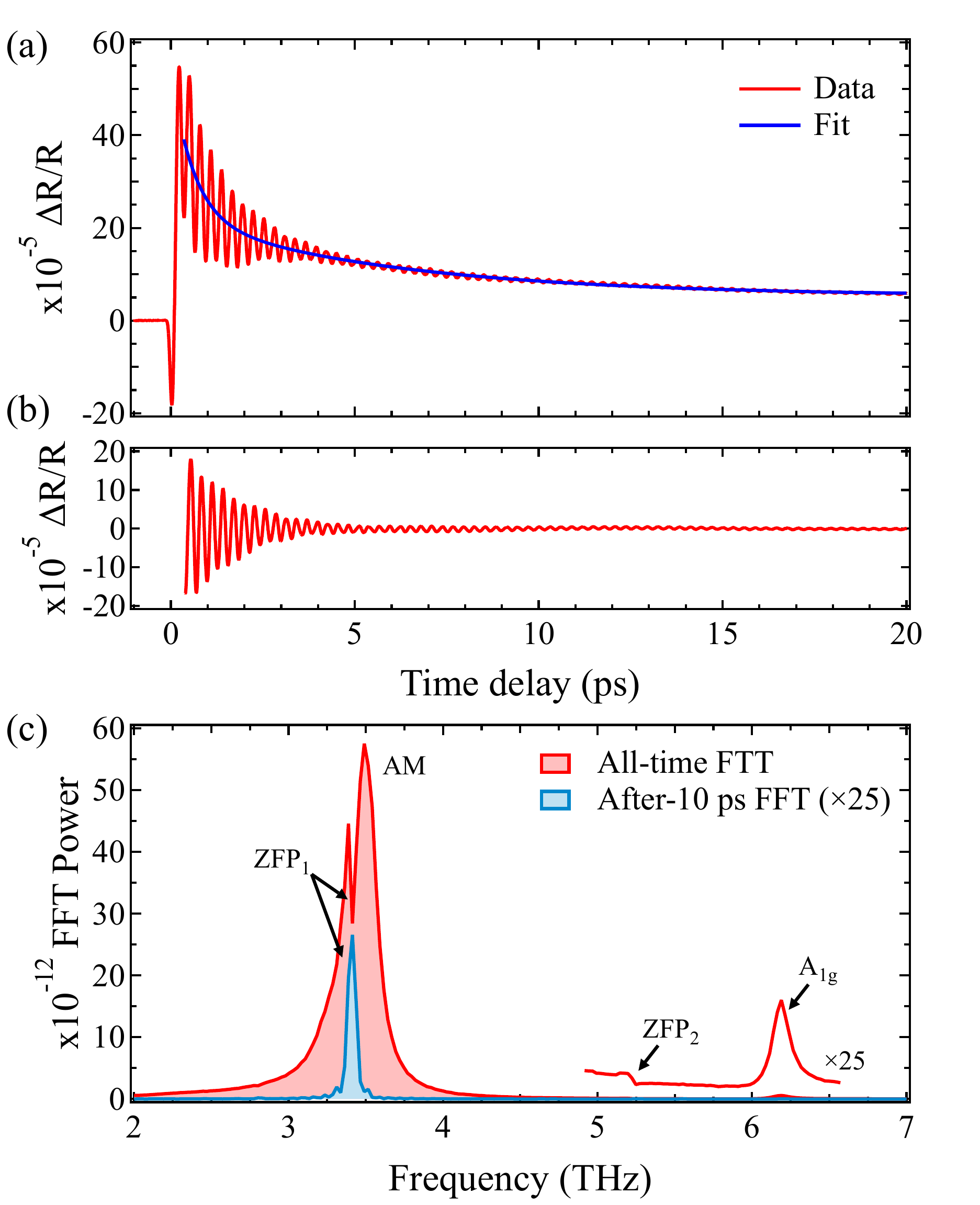}
        \caption{(a) Spectrally integrated $\Delta R/R$ signal at the absorbed pump fluence of 3.45 \textmu J/cm$^2$ under 4.7 K. The blue line represents the fitting result using the sum of two exponential decay and a offset function. (b) The residual component of fitting in (a). Only the coherent oscillations were subtracted. (c) FFT power spectra of the coherent response, where four modes are identified. Blue line represent FFT spectrum after 10 ps singals of (b), where the ZFP$_1$ component was dominant. }
    \label{figure2}
\end{figure}
\clearpage

\begin{figure}[p]
    \centering
    \includegraphics[width = 16cm]{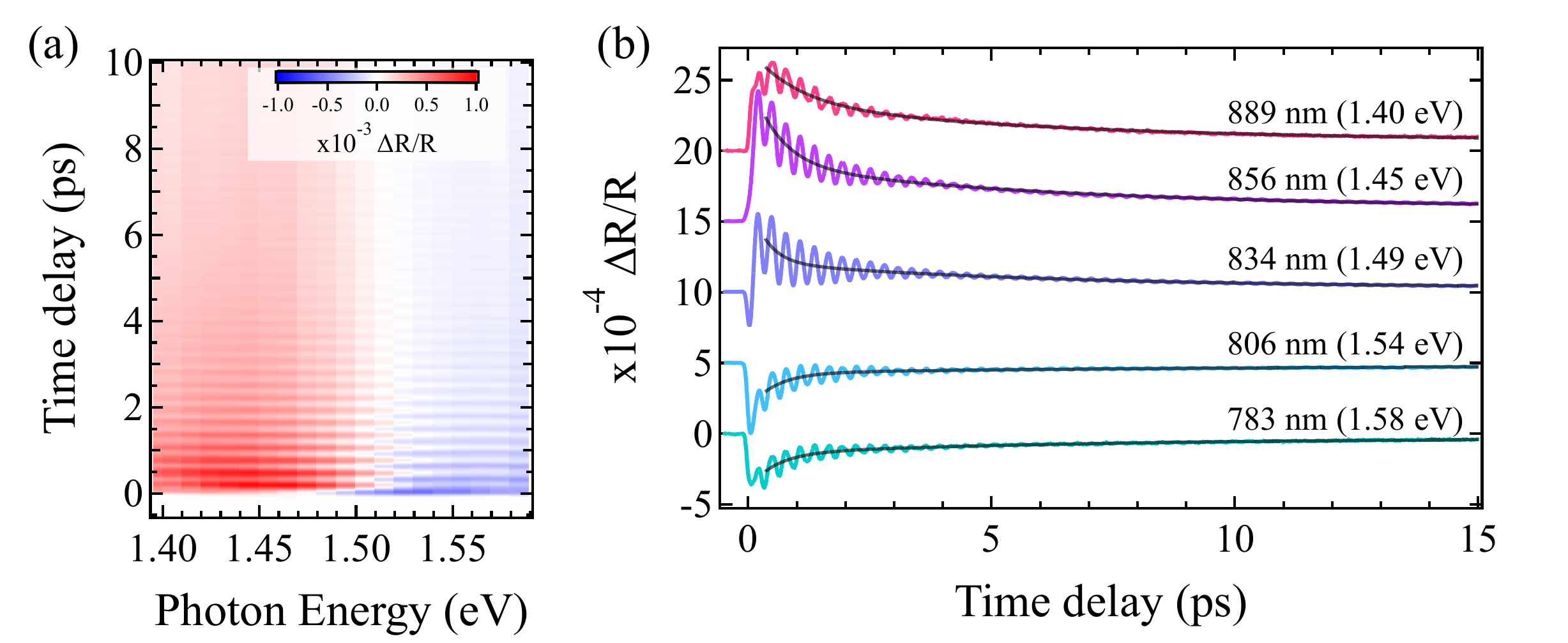}
    \caption{(a) A color map of spectrally resolved $\Delta R/R$ signals. (b) The spectrally resolved $\Delta R/R$ signals at several wavelength. Each trace has vertically offset for clarity. The gray line represents the fitting results using the sum of two exponential decay and a offset function.  % (c) Magnified views of the traces of (b) at several wavelength. The arrows indicate the peak time of each signal.
    }
    \label{figure3}
\end{figure}
\clearpage

\begin{figure}[p]
    \centering
    \includegraphics[width = 16cm]{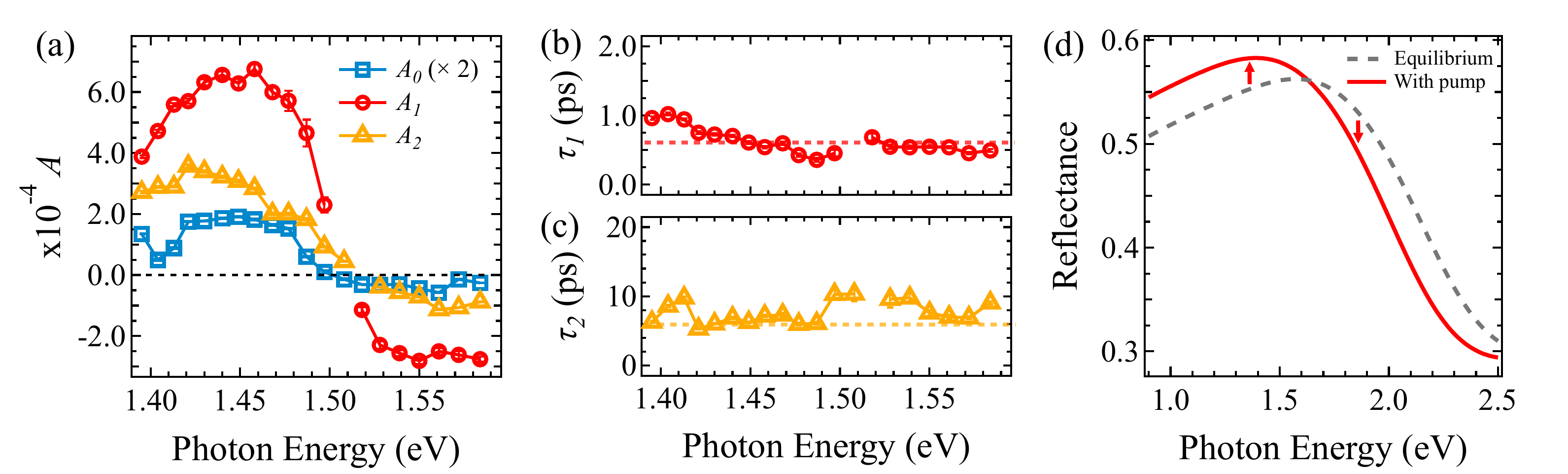}
    \caption{(a) Fitting results for the amplitude $A$ of the exponential and the offset terms.  (b,c) Fitting results for the life time $\tau$ of the exponential components. The dotted lines represent the fitting results of spectrally integrated signal. The error bars in (a), (b), and (c) represent the standard deviations of the fitting. (d) Schematics of the reflectivity change induced by the redshift of the $M$-point optical transition.
    }
    \label{figure4}
\end{figure}
\clearpage

\begin{figure}[p]
    \centering
    \includegraphics[width = 15cm]{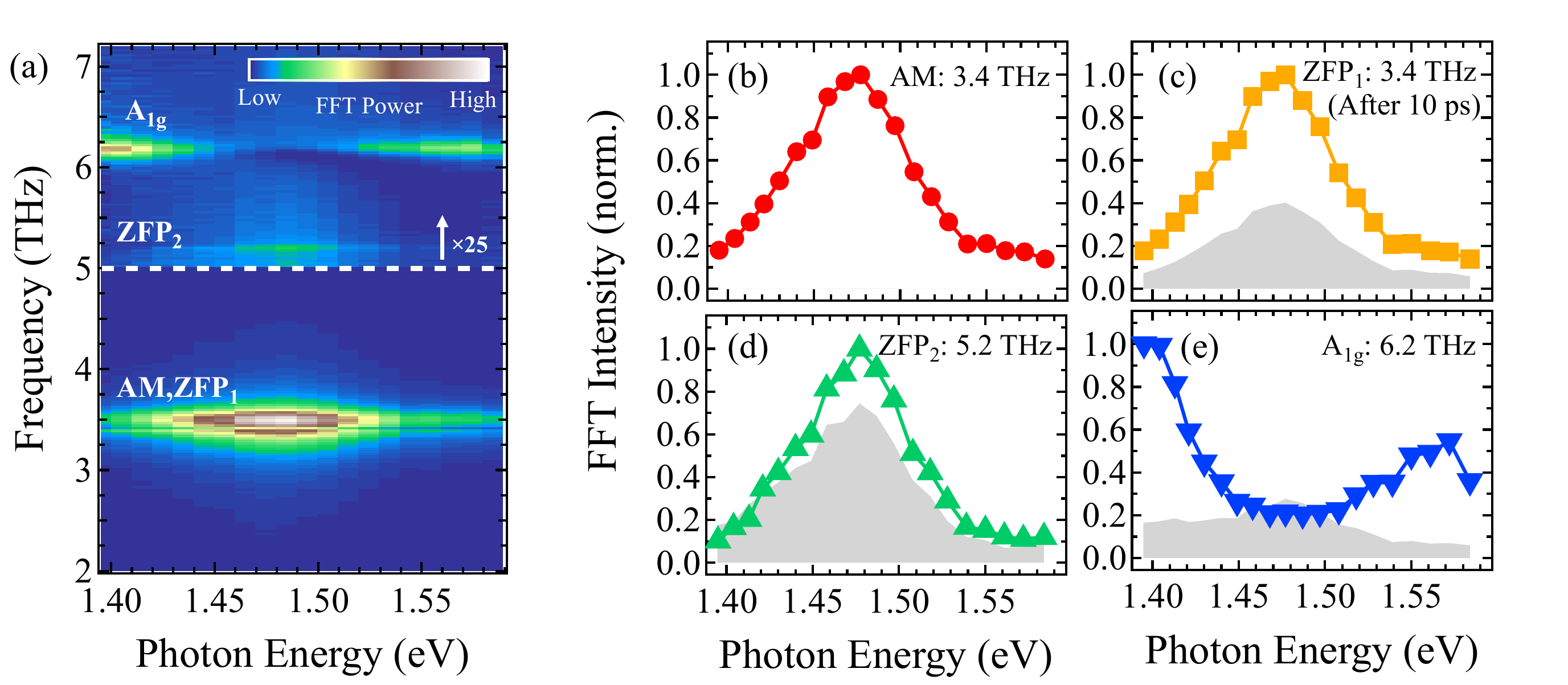}
    \caption{(a) Two-dimensional plot of the FFT power spectrum of the oscillatory components. For visual clarity, the spectrum above 5.0 THz is multiplied by 25. (b-e) FFT intensity of each mode. The gray shaded regions indicate the background in each spectrum. The FFT intensity was obtained by integrating the FFT spectra over each mode, while the background was estimated by integrating the spectra in the surrounding frequency regions. The ZFP$_1$ in (d) was obtained from the FFT after 10 ps region of the $\Delta R/R$.
    }
    \label{figure5}
\end{figure}

\end{document}

\title{Supplemental Material for “Spectroscopic signatures of mode-selective electron-phonon coupling in transient reflectivity change on charge-density-wave TiSe$_2$"}

\author{Yu Mizukoshi}
\affiliation{Department of Applied Physics, Graduate school of Pure and Applied Sciences, University of Tsukuba, 1-1-1 Tennodai, Tsukuba 305-8573, Japan}
\author{Muneaki Hase}
\affiliation{Department of Applied Physics, Graduate school of Pure and Applied Sciences, University of Tsukuba, 1-1-1 Tennodai, Tsukuba 305-8573, Japan}

\maketitle

\section{Additional data for Mode characterization}
In the main text, coherent oscillations at 3.4 THz were assigned to two distinct modes: one was a large-amplitude and short-lived amplitude mode (AM), and the other was a small-amplitude and long-lived zone-folded phonon (ZFP$_1$). In this section, we characterize these modes through temperature-dependent measurements without spectral resolution.

Coherent oscillation parts of transient reflectivity change at each temperature are shown in Fig. S1. The time-domain signal contains a large-amplitude component that decays within 10 ps and a small-amplitude component that remains observable beyond 20 ps. As the temperature increases from 4.8 K, the large-amplitude mode decays more rapidly, whereas the small-amplitude mode remains long-lived. This behavior can also be observed in the FFT spectra shown in Figs. S2(a) and S2(b). While the main peak at 3.4 THz in all-time FFT spectra broadens with increasing temperature, the FFT spectra obtained for the signal after 10 ps exhibit almost no change in either the central frequency or the linewidth. In addition, the dip position at 3.4 THz observed in Fig. S2 (a) remains unchanged even when the main peak undergoes a redshift, and coincides with the peak position in the FFT spectrum after 10 ps [also shown in Fig. 2(c) in the main text], suggesting that this dip originates from interference between the two modes. Therefore, these analyses strongly suggest that the oscillatory components in the 3.4 THz region contain two distinct modes.

For a more detailed analysis, the coherent oscillations were fitted with a sum of damped harmonic oscillations, focusing on the 3.4 THz modes and the 6.2 THz mode that could be reliably evaluated. Figure S3 (a-c) summarizes the fitted parameters. As shown in Fig. S3 (a), the amplitudes $A$ of the 3.40 THz and 3.46 THz modes decrease as the temperature approaches the transition temperature, and they are barely observable at 140 K, suggesting that both modes originate from the CDW transition. Regarding the decay constants $\gamma$ shown in Fig. S3 (b), the short-lived 3.46 THz mode exhibits a faster decay with increasing temperature, while the long-lived 3.40 THz mode shows only a weak temperature dependence, highlighting a clear contrast between the 3.40 THz and 3.46 THz modes. The frequencies $\nu$ shown in Fig. S3(c) also exhibit a clear contrast. The frequency of the short-lived 3.46 THz mode decreases markedly with increasing temperature, whereas that of the long-lived 3.40 THz mode remains nearly unchanged. The temperature dependence of the frequencies agrees well with previous Raman-scattering measurements\cite{sugai1980raman} shown in Fig. S3 (d), supporting the two-mode interpretation.

The fast damping and pronounced softening with increasing temperature observed for the short-lived 3.46 THz mode are characteristic of the AM and have been reported in many CDW materials \cite{sutar2024photo}. For TiSe$_2$, this mode has also been assigned to the AM \cite{snow2003quantum}, whereas the long-lived 3.40 THz mode can be assigned to ZFP$_1$ originating from the CDW superstructure \cite{holy1977raman}.

Overall, these results suggest that two modes contribute to the 3.4 THz region: a short-lived AM at 3.46 THz that decays within 10 ps and a long-lived ZFP$_1$ at 3.40 THz that continues to oscillate beyond 20 ps.

\section{Spectrally-dependent initial response}
In the main text, we noted that the peak-time difference arises from a wavelength-dependent initial response. This interpretation relies on the idea that the peak time of the signal reflects both the rise time of the carrier response and the wavelength-dependent artifact, and variations in the artifact amplitude give rise to the peak-time difference. In this Appendix, we further support this interpretation using additional measurements in which the pulse spectrum was tuned. The central wavelength and spectral bandwidth of the pulse were adjusted by inserting a slit between the prisms inside the femtosecond laser cavity. The upper panel of Fig. \ref{figureAP1}(a) shows the spectrum of the laser pulse used in the main text, while the lower panel shows the tuned spectra. The measurements described in the main text were performed using each tuned laser pulse. The pump fluence and probe wavelength were kept the same within the experimental error.

Figure \ref{figureAP1}(b) shows the $\Delta R/R$ signals obtained with each pulse. Although the probe energies in the respective panels are identical, a pronounced difference is observed in the initial response. For example, when probed at 823 nm (central panel of Fig. \ref{figureAP1}(b)), the red trace, whose spectral center is located on the shorter-wavelength side, shows a weak positive response. In contrast, the blue trace, whose spectral center is located on the longer-wavelength side, exhibits a sharp negative drop. The spectral fingerprints at representative delay times are summarized in Fig. \ref{figureAP1}(c). Around time zero, the response is negative on the higher-energy side of the spectral center indicated by the arrows and positive on the lower-energy side of the spectrum. This feature is consistent with the effect of cross-phase modulation (XPM), which is frequently observed in spectrally resolved measurements \cite{kirby2020transient}.

The idea behind the peak-time difference is as follows. The carrier response is expected to have a rise time of a few hundred femtoseconds governed by electronic thermalization. In the absence of the XPM, the rise time of the signals should be the same. When the XPM is present, however, its contribution can make the apparent signal rise faster. Thus, in the present case, the peak-time difference does not indicate a wavelength-dependent difference in the rise time of the carrier response, but instead originates from the XPM contribution superimposed on the carrier response.

\begin{figure}[p]
    \includegraphics[width = 16cm]{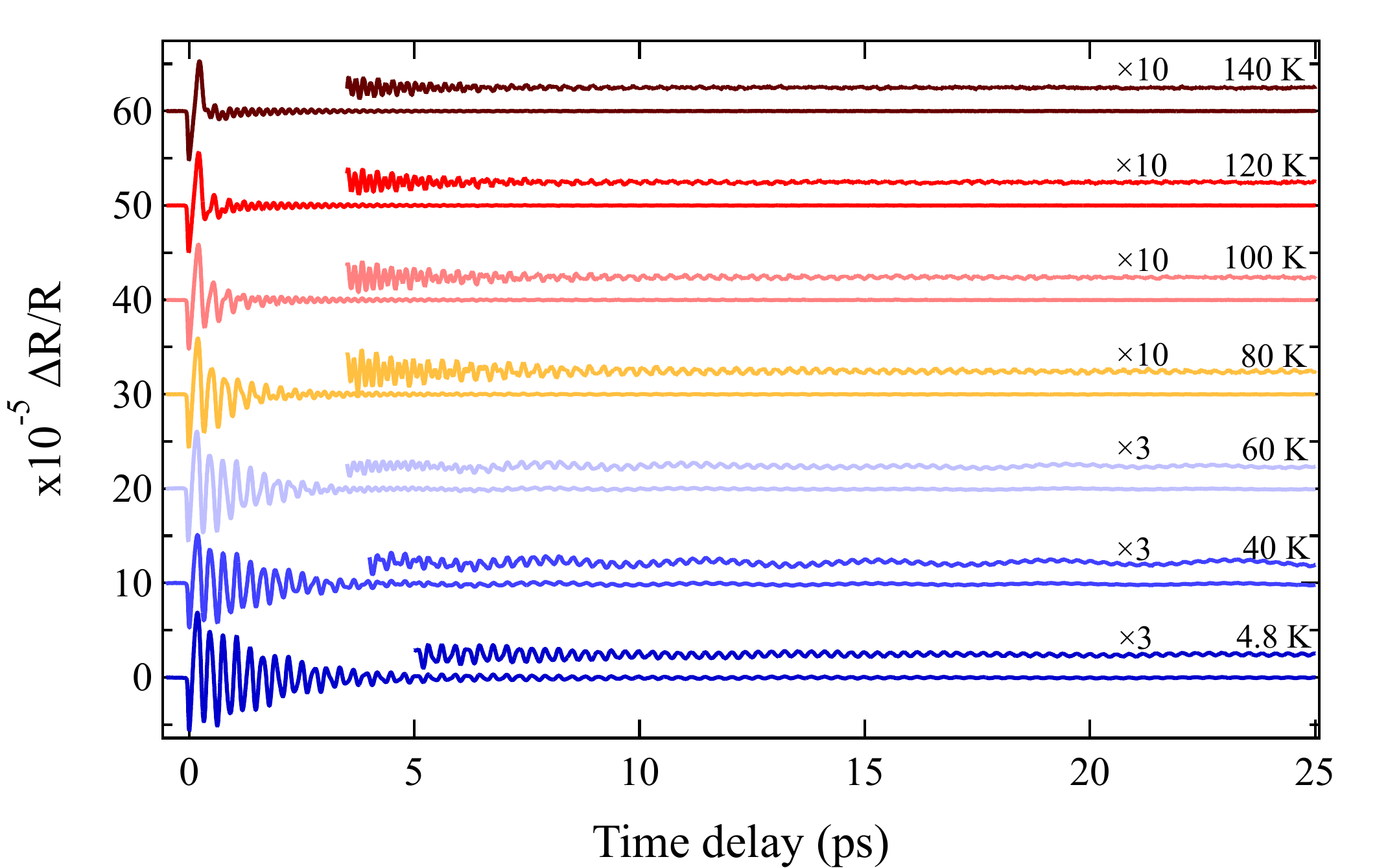}
    \caption{Coherent oscillation part of transient reflectivity change $\Delta$R/R for different temperatures. The pump fluence was fixed at 5 \textmu J/cm$^2$.}
    \label{FigS1}
\end{figure}

\begin{figure}[p]
    \includegraphics[width = 16cm]{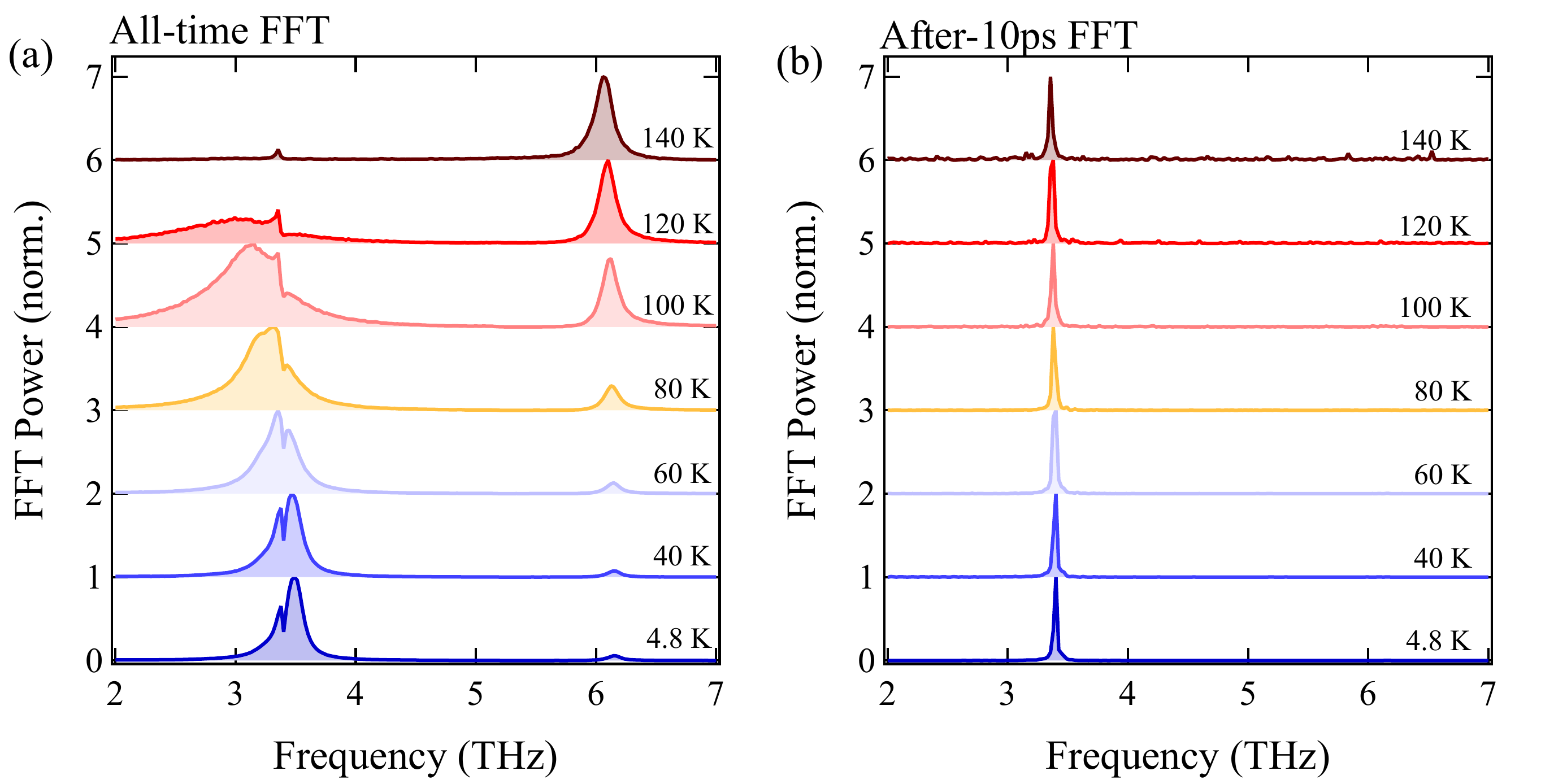}
    \caption{(a,b) FFT power spectra for $\Delta$R/R signals including all-time data and data after 10 ps, respectively.}
    \label{FigS2}
\end{figure}

\begin{figure}[p]
    \includegraphics[width = 12cm]{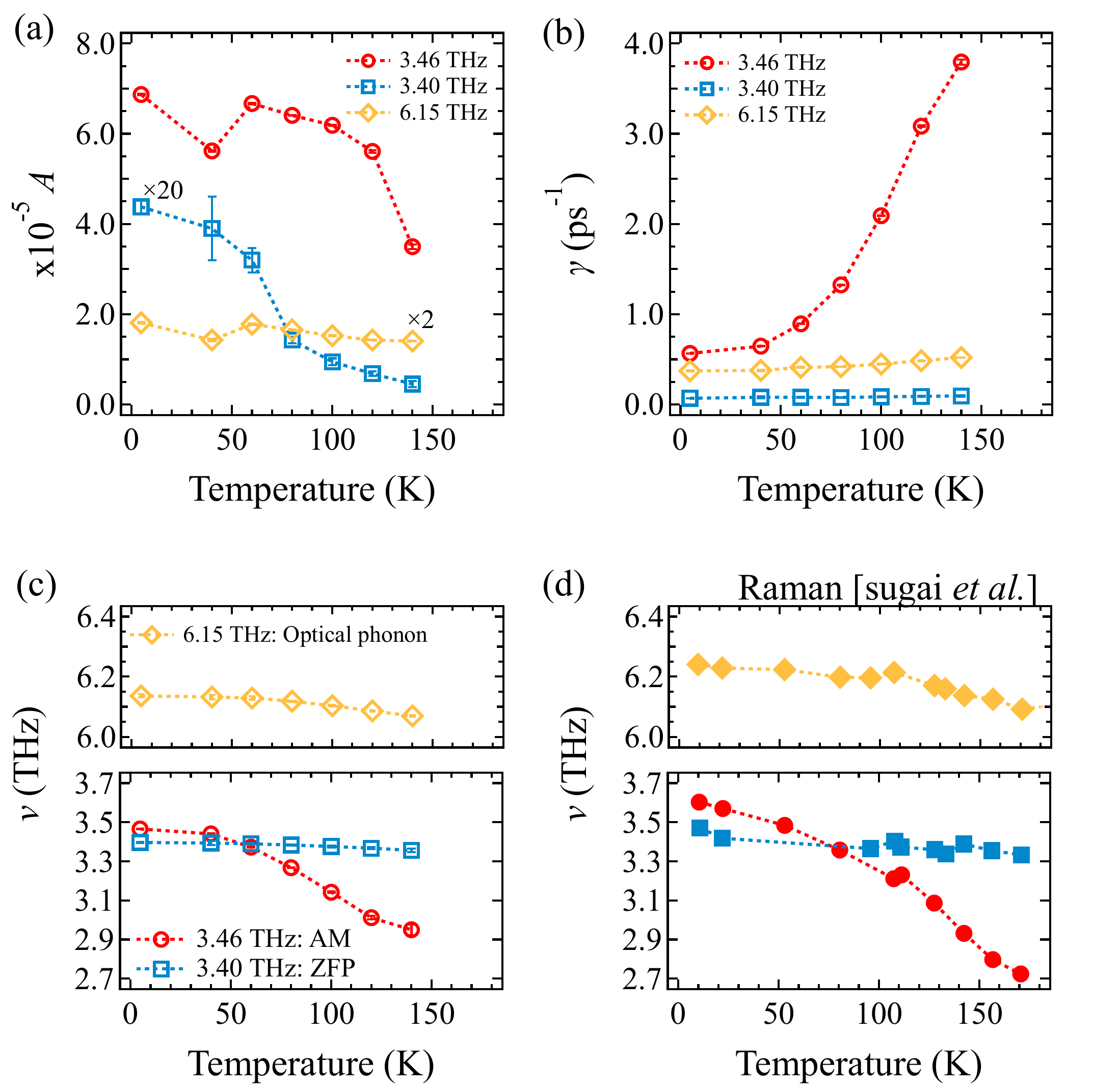}
    \caption{(a-c) Fitting result parameters obtained by damped harmonic oscillation function. (d) Temperature-dependent phonon frequencies reported by Raman scattering \cite{sugai1980raman}. The error bars represent the standard deviations of the fitting parameters, although most of them are within the symbols.
    }
    \label{FigS3}
\end{figure}

\begin{figure}[p]
    \centering
    \includegraphics[width = 16.2cm]{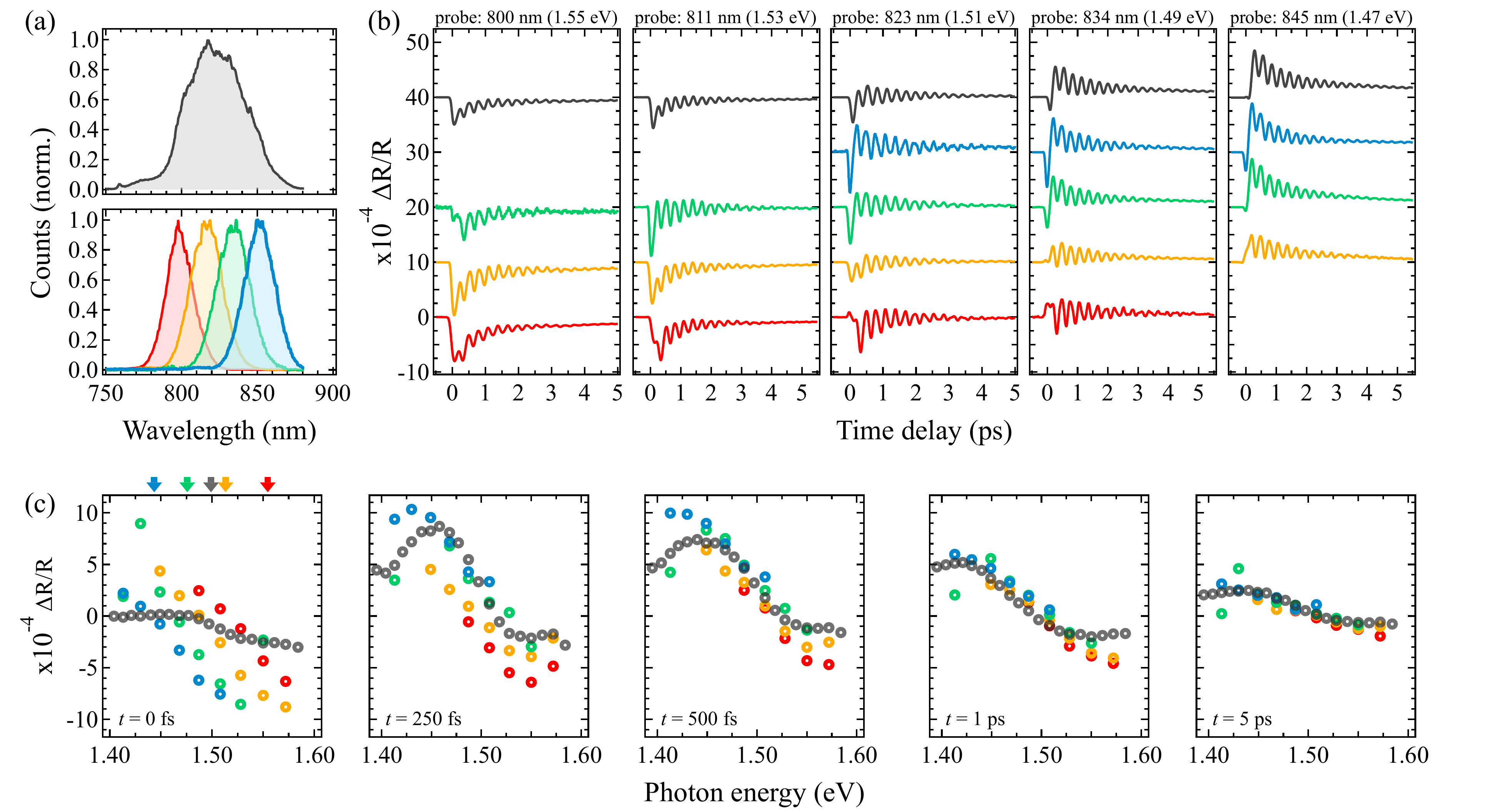}
    \caption{(a) The spectrum used in the main text (upper panel), and wavelength-tuned pulse spectra, (lower panel). (b) Probe-wavelength-resolved $\Delta R/R$ signals. Each panel corresponds to the same probe wavelength, and the fundamental wavelengths of the pulses are represented by the same colors as those used in (a). (c) $\Delta R/R$ signals at selected delay times. The arrows in the left panel indicate the spectral centers of the pulses.
    }
    \label{figureAP1}
\end{figure}
\clearpage

\bibliography{reference}% Produces the bibliography via BibTeX.